
\input harvmac
\noblackbox

\def\eg{{\it e.g.}}
\def\ie{{\it i.e.}}

\lref\Antoniadis{I. Antoniadis and T. R. Taylor, Nucl. Phys. {\bf
B731} (2005) 164, hep-th/0509048.}

\lref\DT{A. Animisov, M. Dine, M. Graesser and S. Thomas, Phys. Rev.
{\bf D65} (2002) 105011, hep-th/0111235; JHEP {\bf 0203} (2002) 036,
hep-th/0201256.}

\lref\bdthm{T. Banks and L. Dixon,
 Nucl. Phys. {\bf B307}, 93 (1988).}

\lref\dsww{M. Dine, N. Seiberg, X.G. Wen and E. Witten, Nucl. Phys.
{\bf B278} (1986) 769.}

\lref\Savas{S. Dimopoulos, S. Raby and F. Wilczek, Phys. Rev. {\bf
D24} (1981) 1681.}

\lref\bkq{C.P. Burgess, R. Kallosh and F. Quevedo, JHEP {\bf 0310}
(2003) 056, hep-th/0309187.}

\lref\DougAlb{I. Brunner, M. Douglas, A. Lawrence and C. Romelsberger,
JHEP {\bf 0008} (2000) 015, hep-th/9906200.}

\lref\Herman{H. Verlinde and M. Wijnholt, hep-th/0508089.}

\lref\witteno{E. Witten, in {\it Shelter~Island~II}, 1983.}

\lref\warpedno{ G. Gibbons, in {\it GIFT Seminar~1984}; J. Maldacena
and C. Nunez, Int J. Mod. Phys. {\bf A16} (2001) 822,
hep-th/0007018.}

\lref\EvadS{E. Silverstein, hep-th/0106209.}

\lref\EdG{E. Witten, Nucl. Phys. {\bf B471} (1996) 135,
hep-th/9602070.}

\lref\Cumrun{S. Gukov, H. Ooguri and C. Vafa, unpublished.}

\lref\Sharpe{E. Sharpe, ATMP {\bf 2} (1998) 1441, hep-th/9810064.}

\lref\RS{L. Randall and R. Sundrum, Nucl. Phys. {\bf B557} (1999)
79, hep-th/9810155.}

\lref\GM{D.E. Diaconescu, B. Florea, S. Kachru and P. Svr\v{c}ek,
hep-th/0512170.}

\lref\Murayama{G. Giudice, M. Luty, H. Murayama and R. Rattazzi,
JHEP {\bf 9812} (1998) 027, hep-ph/9810442.}

\lref\Burgess{C.P. Burgess, C. Escoda and F. Quevedo,
hep-th/0510213.}

\lref\Oliver{O. DeWolfe and S.B. Giddings, Phys. Rev. {\bf D67}
(2003) 066008, hep-th/0208123.}

\lref\Gaugino{ D.E. Kaplan, G. Kribs and M. Schmaltz, Phys. Rev.
{\bf D62} (2000) 0350, hep-ph/9911293; Z. Chacko, M. Luty, A.E.
Nelson and E. Ponton, JHEP {\bf 0001} (2000) 003, hep-ph/9911323.}

\lref\largef{T. Banks, M. Dine, P.J. Fox and E. Gorbatov, JCAP {\bf
0306} (2003) 001, hep-th/0303252.}

\lref\DGKT{O. DeWolfe, A. Giryavets, S. Kachru and W. Taylor, JHEP
{\bf 0507} (2005) 066, hep-th/0505160.}

\lref\Zwirner{G. Villadoro and F. Zwirner, JHEP {\bf 0506} (2005)
047, hep-th/0503169; P.G. Camara, A. Font and L.E. Ibanez, JHEP {\bf
0509} (2005) 013, hep-th/0506066.}

\lref\vafa{C. Vafa, hep-th/0509212.}

\lref\Dine{M. Dine, R. Leigh and A. Kagan, Phys. Rev. {\bf D48}
(1993) 4269, hep-ph/9304299.}

\lref\LutySun{M. Luty and R. Sundrum, Phys. Rev. {\bf D62} (2000)
035008, hep-th/9910202.}

\lref\kha{V. Balasubramanian and P. Berglund, JHEP {\bf 0411} (2004)
085, hep-th/0408054.}

\lref\khb{V. Balasubramanian, P. Berglund, J. P. Conlon and F.
Quevedo, JHEP {\bf 0503} (2005) 007, hep-th/0502058.}

\lref\khc{G. von Gersdorff and A. Hebecker, Phys. Lett. {\bf B624}
(2005) 270, hep-th/0507131.}

\lref\Haack{M. Berg, M. Haack and B. K\"ors, hep-th/0508171.}

\lref\bbhl{M. Becker, K. Becker, M. Haack and J. Louis, JHEP {\bf
0206} (2002) 060, hep-th/0204254; M. Berg and M. Haack, Phys. Rev.
{\bf D71} (2005) 026005, hep-th/0404087; M. Berg, M. Haack and B.
K\"ors, JHEP {\bf 0511} (2005) 030, hep-th/0508043. }

\lref\delmoral{M. Garcia del Moral, hep-th/0506116.}

\lref\extranatural{N. Arkani-Hamed, H.C. Cheng, P. Creminelli and L.
Randall, Phys. Rev. Lett. {\bf 90} (2003) 221302, hep-th/0301218.}

\lref\nima{N. Arkani-Hamed, L. Motl, A. Nicolis and C. Vafa,
hep-th/0601001.}

\lref\dixon{L. Dixon, V. Kaplunovsky and C. Vafa, Nucl. Phys. {\bf
B294} (1987) 43.}

\lref\susskind{L. Susskind, hep-th/0302219.}

\lref\moore{J. Horne and G. Moore, Nucl. Phys. {\bf B432} (1994)
109, hep-th/9403058.}

\lref\mike{M. Douglas, talk at Strings 2005; M. Douglas and Z. Lu,
hep-th/0509224.}

\lref\nflation{S. Dimopoulos, S. Kachru, J. McGreevy and J.G.
Wacker, hep-th/0507205.}

\lref\Pomarol{A. Pomarol and R. Rattazzi, JHEP {\bf 9905} (1999)
013, hep-ph/9903448;
Z. Chacko, M. Luty, I. Maksymyk and E. Ponton, JHEP {\bf 0004} (2000) 001,
hep-ph/9905390;
E. Katz, Y. Shadmi and Y. Shirman, JHEP {\bf 9908} (1999) 015,
hep-ph/9906296.
}

\lref\Eva{E. Silverstein, hep-th/0405068.}

\lref\gsw{See M.B. Green, J. Schwarz and E. Witten, {\it Superstring
Theory}, Volume II, Cambridge University Press (1987), pp. 456-457.}

\lref\liam{R. Easther and L. McAllister, hep-th/0512102.}

\lref\dvalitye{G. Dvali and S.H. Tye, Phys. Lett. {\bf B450} (1999)
72, hep-ph/9812483.}

\lref\kklt{S. Kachru, R. Kallosh, A. Linde and S. Trivedi, Phys.
Rev. {\bf D68} (2003) 046005, hep-th/0301240.}

\lref\kklmmt{S. Kachru et al, JCAP {\bf 0310} (2003) 013,
hep-th/0308055.}

\lref\giddings{S.B. Giddings, Phys. Rev. {\bf D68} (2003) 026006,
hep-th/0303031.}

\lref\lyth{D. Lyth, Phys. Rev. Lett. {\bf 78} (1997) 1861,
hep-ph/9606387; G. Efstathiou and K. Mack, JCAP {\bf 0505} (2005)
008, astro-ph/0503360.}

\lref\Witten{E. Witten, Nucl. Phys. {\bf B474} (1996) 343,
hep-th/9604030.}

\lref\IIBflux{S. Gukov, C. Vafa and E. Witten, Nucl. Phys. {\bf
B584} (2000) 69, hep-th/9906070; K. Dasgupta, G. Rajesh and S.
Sethi, JHEP {\bf 9908} (1999) 023, hep-th/9908088.}

\lref\GKP{ S.B. Giddings, S. Kachru and J. Polchinski, Phys. Rev.
{\bf D66} (2002) 106006, hep-th/0105097.}

\lref\theshining{
  N.~Arkani-Hamed and S.~Dimopoulos,
  Phys.\ Rev.\  {\bf D65} (2002) 052003,
 hep-ph/9811353.
}

\Title{\vbox{\baselineskip12pt \hbox{hep-th/0601111}
\hbox{SU-ITP-06/02} \hbox{SLAC-PUB-11618} }}
{\vbox{\centerline{Bounds on masses of bulk fields}
\vskip2pt\centerline{in string compactifications}}}

\centerline{Shamit Kachru,\footnote{$^1$}{skachru at stanford.edu}
John McGreevy\footnote{$^2$}{mcgreevy at stanford.edu}
and Peter Svr\v{c}ek\footnote{$^3$}{svrcek at stanford.edu}
}
\medskip\centerline{Department of Physics and SLAC,
Stanford University} \centerline{Stanford, CA 94305/94309, USA}

\vskip .15in In string compactification on a manifold $X$, in
addition to the string scale and the normal scales of low-energy
particle physics, there is a Kaluza-Klein scale ${1\over R}$
associated with the size of $X$. We present an argument that generic
string models with low-energy supersymmetry have, after moduli
stabilization, bulk fields with masses which are parametrically
lighter than ${1\over R}$. We discuss the implications of these
light states for anomaly mediation and gaugino mediation scenarios.

\Date{January 2006}

\newsec{Introduction}

It is interesting to ask what constraints string theory can put on
models of low-energy physics.  In the framework of worldsheet
conformal field theory, some powerful results were derived in the
1980s. For instance, one could bound the rank of any non-abelian
factors in the 4d gauge group to be $\leq 22$, one could prove that
type II theories were incapable of incorporating the field content
of the Standard Model \dixon,  and one could make a precise argument
substantiating the long-standing belief that quantum gravity does
not allow continuous global symmetries \bdthm.

Of course, these results provide a cautionary tale; the advances of
the duality revolution extended our understanding of string theory
and added ingredients which allow one to circumvent the first two
``no-go'' theorems just described. Nevertheless, the general program
of outlining what is and is not expected to be possible in
string-derived effective field theories remains intriguing. It seems
quite clear that ${\it some}$ 4d effective field theories, coupled
to gravity, cannot be UV completed by string theory. The real
questions are, what are the precise string theoretic constraints on
effective field theories? And, perhaps more importantly, do any
of these constraints impinge on effective field theories that have
been seriously proposed as models of the world?

In the past few years, there have been many results in this area.
Several of these results apply to string cosmology. For instance,
there are arguments that, under reasonable assumptions, string
theory models of de Sitter space are always metastable with a
lifetime short compared to the recurrence time
\refs{\kklt,\susskind,\giddings}.\foot{An early discussion of the
possible importance of metastability in addressing conceptual
challenges of de Sitter quantum gravity appeared in \EvadS.}
In
addition, seemingly natural models of large-field inflation based on
super-Planckian axion decay constants \extranatural\ may not exist
in string theory \largef.\ More elaborate constructions may allow
analogues of chaotic inflation in string theory
\refs{\nflation,\liam}; observation of a tensor-to-scalar ratio $r
\geq .05$ by the next generation of experiments would make this a
central issue \lyth. Other popular models of inflation based on
brane motion in extra dimensions \dvalitye\ were similarly
constrained by genericity arguments in \kklmmt.

There are also interesting results which constrain the geometry of
stringy moduli spaces of vacua \refs{\moore,\mike,\vafa}. Most
recently, there has been a striking claim that theories with abelian
gauge fields coupled to quantum gravity, with perturbative gauge
coupling $g$, must manifest a new physical scale $\sim g M_P$ to
avoid the paradoxes associated with remnants \nima.

In this note, we provide simple arguments that indicate that in
string compactification on a manifold $X$ of radius $R$ preserving
low-energy supersymmetry, there will generically be bulk modes of
mass parametrically $< {1\over R}$. This is relevant in evaluating
the prospects for phenomenological scenarios like anomaly mediation
\refs{\RS,\Murayama} and gaugino mediation \Gaugino\ to be UV
completed by string theory.\foot{Other discussions of possible
difficulties in implementing anomaly mediation in string theory
appear in \refs{\DT,\Antoniadis}.} As these scenarios constitute
some of the leading ideas for solving the SUSY flavor problem, our
result may be relevant in understanding which classes of realistic
SUSY models emerge from string theory.

There may be other interesting applications for such a bound.
As a simple example, we mention
other attempts at sequestering symmetry breaking
in extra dimensions, such as
(supersymmetric implementations of)
``shining" of flavor symmetry breaking \theshining.
In such models, the hierarchy between Standard Model Yukawa couplings
is provided by
the Yukawa-suppressed couplings between
the Standard Model and some distant
${\cal O}(1)$ source of flavor violation.

In \S2, we review why, in anomaly and gaugino mediation, one wishes to find
compactifications where all bulk fields have masses $\geq {1\over
R}$.  In \S3, we show that known ways of stabilizing the moduli of
$X$ using the ingredients in 10d string theory will generically
produce masses which are parametrically $\ll {1 \over R}$ in the
regime $ R \gg l_s$ where one can trust classical geometry. It
will be clear that
the arguments can be extended to constrain other higher-dimensional
theories which do not manifest significant non-locality at distances
larger than $R$.

Our arguments work most simply for models where the
extra dimensions are reasonably symmetric (\ie\ characterized by a
single scale $R$).
We discuss asymmetric models in \S4\ (focusing on heterotic M-theory
for concreteness).
In such models, say with two characteristic
radii $R >> R_X$,
one can
imagine two different conjectures.
A strong conjecture would posit that there are always masses
parametrically $< 1/R$.  A weak conjecture would instead be
that there are always
masses parametrically $< 1/R_X$.  In some circumstances,
the success of the extra-dimensional
mediation mechanisms may only require violations of the strong
conjecture, or may even be consistent with the strong conjecture if
the light fields have special constraints on their couplings.
We find that both conjectures hold in commonly studied models.
For the strong conjecture,
one can find regimes where
the question becomes numerical instead of parametric for a subset of
the moduli fields (while others remain parametrically lighter than $1/R$).
If the numerical factors conspire properly for the moduli whose
masses are $\sim 1/R$, and
if special considerations allow one to argue that the remaining
parametrically lighter moduli have constrained (flavor-blind)
couplings to the Standard Model (as sometimes happens \LutySun), one
could potentially make a working model of anomaly mediation.
We close with a brief
discussion in \S5.

\newsec{Solving the flavor problem with bulk locality}

The MSSM and its extended counterparts typically introduce many new
opportunities for flavor violation into low-energy physics.  For
instance, if the soft-breaking terms include a generic squark mass
matrix, the resulting flavor violation is unacceptable. For one
discussion of this issue with further references, see \Dine.

Several natural mechanisms have been proposed to surmount this
problem. The basic idea is to ensure that the messenger of SUSY
breaking from some hidden sector to the Standard Model, couples to
squarks in a flavor-blind way.  The gauge interactions of the
Standard Model itself clearly do this. So one promising mechanism
which can surmount the flavor problem is gauge mediation, where SUSY
breaking is first transmitted to the Standard Model gauge
multiplets, whose gauge couplings to matter particles then lead to
squark and slepton masses. It seems plausible that pseudo-realistic
models of gauge mediation can be naturally embedded into string
theory \GM.

Two other very interesting proposals for solving the flavor problem,
anomaly mediation and gaugino mediation, use the physics of extra
dimensions. In anomaly mediation \refs{\RS, \Murayama}, the fact
that the ``conformal compensator'' multiplet of supergravity couples
universally to mass dimension, can be used to impart universal
masses to sparticles if the dominant F-term arises in the conformal
supermultiplet. The couplings of this field to squarks and sleptons
are Planck suppressed, and for the flavor-blind contribution of
anomaly mediation to dominate over gravity mediation, one needs to
${\it sequester}$ away the SUSY breaking sector.  That is, if $Z$ is
the chiral superfield with the dominant F-term, one needs to forbid
non-universal, radiatively induced terms of the form
\eqn\badterms{\int d^4 \theta~ c_i {1\over M^2} Z^\dagger Z
Q_i^\dagger Q_i ,} where $M$ is the scale of high-energy physics
that has been integrated out in writing down the effective field
theory (typically it could be the Planck scale or the string scale),
$Q_i$ are Standard Model particles, and $c_i$ are ${\cal O}(1)$
coefficients. Terms with the structure \badterms\ are generically
induced by integrating out high-scale physics.

In order to forbid such terms, it was suggested in \RS\ that one
should find models where the SUSY-breaking hidden sector and the
Standard Model arise on branes which are well-separated in the
compactification manifold $X$ of radius $R \gg {1\over M}$. Suppose
they are separated by a distance $d \leq R$. Assume furthermore that
all bulk moduli fields $\phi_a$ have masses \eqn\massbound{m_a
> {1\over d}~.} The couplings between the field $Z$, localized on
the hidden brane, and the fields $Q_i$, localized on ``our'' brane,
which are generated by exchange of the bulk field $\phi_a$ will then
be suppressed by the factor \eqn\falloff{c_i \sim e^{-m_a d} \ll 1}
if \massbound\ is satisfied. The dangerous flavor-violating effects
can therefore be suppressed.\foot{The reader may wonder about the
possible effects of Kaluza-Klein modes.  It is argued in \RS\ and
in more detail in \LutySun\ that
integrating these out does not generate the dangerous operators.}

Anomaly mediation, by itself, leads to tachyonic sleptons. With some
cleverness, reasonable extensions of the minimal scenario can be
constructed that avoid this problem.  Specific extensions
are discussed in, for instance, \Pomarol.

Precisely the same idea of sequestering enables models of gaugino
mediation \Gaugino\ to solve the flavor problem. In such models,
SUSY is again broken on a distant hidden sector brane. However, in
these scenarios, the Standard Model gauge fields and gauginos live
in (at least part of) the bulk of $X$. The gauge multiplet couples
directly to the hidden brane, which yields a gaugino mass. This then
generates flavor-blind Standard Model splittings as in gauge
mediation. One finds that the sparticle masses are suppressed by
powers of $1/M$ in a way that depends on the number of extra
dimensions where the gaugino lives. Again, for the effects of
gaugino mediation to dominate bulk exchange, the regime \massbound\
for masses of bulk moduli is favored.

In the following sections, we argue that achieving the bound
\massbound\ for bulk moduli masses in string theory is difficult. More
precisely, the scaling of all known effects which can generate a
potential for bulk moduli, leave some masses which are parametrically
smaller than \massbound\ in the well-studied classes of
supersymmetric string models.
We will see that this result follows from
the simple fact that string models give rise to local effective theories at
distances large compared to the string scale.

\newsec{The scaling of energy densities in string theory}

Let us assume that the bulk geometry is well described by a 6d
compact manifold $X$ which preserves supersymmetry at the KK scale
${1\over R}$. In our discussion we will assume $X$ is a Calabi-Yau
space but relaxing this assumption would not modify our conclusions.

The 10d metric takes the form \eqn\metric{ds^2 = e^{2A(y)}
\eta_{\mu\nu} dx^\mu dx^\nu + R^{2} \tilde g_{mn}(y) dy^m dy^n}
where $y^m$ are coordinates on $X$ and $R$ is the overall radial
dilaton. Here $A(y)$ is a possible warp factor.  The scenarios of
interest do not rely on a heavily warped geometry and we will
henceforth set ${e^A} \sim 1$.

The field content of the theory depends on our precise choice of
string theory.  For definiteness we will mostly consider the wide
class of IIB Calabi-Yau orientifolds whose effective field theories
were derived in \refs{\GKP,\Oliver}. It will be clear that our
arguments could be generalized to other known classes of geometric
compactifications with minimal modifications.  We make some comments
about another specific class of models in \S4.

In IIB string theory on a Calabi-Yau space $X$, the relevant fields
include a dilaton-axion, the K\"ahler and complex structure moduli
of $X$, and two three-form field strengths $H_{NS}$ and $F_{RR}$.
There is also a five-form field strength $F_5$ which is self-dual;
since $X$ has no nontrivial five-cycles it will play no role in our
discussion.

Our question will be, how does the potential for the bulk moduli
(the axio-dilaton, the K\"ahler moduli, and the complex structure
moduli) scale with $R$? The following discussion is very similar to
the one in \giddings, where the focus was on the potential for the
radial dilaton. A nice pedagogical description of similar results
appears in \Eva.

The basic point is the following.  Below the scale $1/R$ there is a
4d supersymmetric effective field theory.  The scalar potential for
bulk moduli \eqn\potis{V = e^{{K \over M_P^2}} \left( \sum_{i,j}
g^{i \bar j} D_i W D_{\bar j} \overline{W} - 3 {|W|^2 \over M_P^2}
\right) ~} can be computed from a superpotential $W$ and a K\"ahler
potential $K$.  The finite list of ingredients in 10d supergravity,
together with the known possible stringy corrections to the
superpotential and the known expansion parameters which control
corrections to $K$, allow us to categorize the possible scalings of
$V$ with $1/R$. We will see that the masses of bulk moduli end up
parametrically below the scale $1/R$.  Of course this was necessary
to have a 4d effective theory incorporating the moduli at all (since
one could not self-consistently keep fields with $m \sim 1/R$
without keeping the KK tower as well).  We will set $g_s \sim 1$ for
this discussion, but let us note that in models with weak string
coupling, there could be further suppression of the bulk moduli
masses by powers of $g_s$.

\subsec{Potentials from bulk fluxes}

Turning on magnetic fluxes in the Calabi-Yau space $X$ generates a
potential for many of the moduli fields.  Such models have been
extensively studied, some representative references include
\refs{\IIBflux,\GKP}.

The scaling of the potential arising from the 3-form fluxes can be found as
follows. Start with the 10d Lagrangian \eqn\tend{\int d^{10}x
~\sqrt{-g_{10}} \left(F_{ijk}g_{10}^{il} g_{10}^{jm} g_{10}^{kn}
F_{lmn} + \cdots\right)} The reduction to 4d using the ansatz
\metric\ is straightforward. The three factors of $g^{..}$ give a
${1\over R^6}$.  The volume of $X$ gives a compensating factor of
$R^6$.  This naively gives an overall $R^0$ scaling.

However, it is important to remember that one should measure
energies in 4d Einstein frame.  The naive reduction gives an
Einstein term with an $R^6$ in front.  Doing the Weyl rescaling of
the four dimensional metric by $1/R^6$ to reach the Einstein frame
and re-writing the potential $\sqrt{-g_4} V$ in terms of the
Einstein metric, yields a factor of $R^{-12}$.

Hence, the overall energy from the three-form fluxes scales like
${1\over R^{12}}$.  To find the scale of the moduli masses $m$ which
result from this potential, one equates \eqn\modulimass{M_P^2 m^2
\sim M_P^4 \left({\alpha^\prime \over {R^2}} \right)^{6}~.} Recall
that \eqn\mp{M_P^2 \sim {R^6 \over {(\alpha^\prime)^4}}~.} We
conclude that the moduli which receive masses from three-form fluxes
in string theory have \eqn\fluxsca{m \sim {\alpha^\prime \over
R^3}~.} This is parametrically $\ll {1\over R}$ at large $R$. For a
more detailed analysis which reaches the same conclusion, see \eg\
\S5.4 of \Oliver.

A different way to derive the moduli mass \fluxsca\ that does not
use the Weyl rescaling is to find the Lagrangian for the moduli and
read off the mass from it. The bulk moduli in question, the complex
and K\"ahler structure moduli, are fluctuations of the metric. Hence
their kinetic term comes from expanding the GR action $\int_{X\times
R^4} \sqrt{g_{10}} R$ in small fluctuations. Reducing this to four
dimensions, the kinetic term gets a factor of $M_P^2$ in front
\eqn\modure{L_{{\rm kinetic}} (\phi) \sim M_P^2 (\partial\phi)^2,}
just like the 4d GR action, which is essentially the kinetic term
for fluctuations of the 4d metric. Let us assume that the moduli
potential from the fluxes has scale $V_0.$ Its form near a local
minimum, assuming a generic shape of the potential, is
\eqn\potential{V(\phi)\sim V_0 \phi^2,} since $V\sim V_0$ for
$\phi\sim 1$. Normalizing $\phi\rightarrow \phi/M_P$ to get
canonical kinetic terms yields a general formula for the moduli mass
\eqn\massm{m_\phi\sim {\sqrt{V_0}\over M_P}.} The complex structure
moduli considered above have $V_0\sim (\alpha^\prime)^{-2}$ and
$M_P^2\sim R^6/(\alpha^\prime)^4,$ so that $m_\phi\sim \alpha'/R^3,$
in agreement with \modulimass.\

\subsec{Potentials from brane worldvolume fluxes}

Many pseudo-realistic string models incorporate D-branes wrapping
cycles in $X$.  D-branes come equipped with worldvolume gauge
fields.  So, for instance, in type IIB string theory, in addition to
the bulk three-form field strengths one can in general also consider
two-form field strengths of the brane worldvolume gauge fields.  In
the class of type IIB compactifications described in \GKP, the
unique possibility consistent with Lorentz invariance is nontrivial
gauge field strengths supported on D7 brane worldvolumes. The
effects of these fluxes have been discussed extensively in recent
literature, see for instance \refs{\bkq,\delmoral}.

The 4d description of the potential arising from D7 brane gauge
fields is somewhat complicated, but if one is only interested in
scalings as we are, the result is rather simple.  The energy stored
in a worldvolume gauge field $F$, on a D7 brane wrapping the 4-cycle
$C$, is \eqn\Fen{E \sim \int d^4x \int_{C} \sqrt{-g_8}
F_{mn}g^{mq}g^{nr}F_{qr}~.} After the reduction to four dimensions
and Weyl rescaling to Einstein frame, this contribution to the 4d
potential scales as ${1/ R^{12}}$, similarly the contribution from
the bulk fluxes described in \S3.1. Hence, the mass of the moduli
generated by D7-brane worldvolume fluxes is \eqn\dscaling{m\sim
{\alpha^\prime\over R^3}~.}

\subsec{Quantum corrections to $W$}

The flux superpotential in IIB Calabi-Yau models does not depend on
the K\"ahler moduli of $X$.  However, there are model-dependent
nonperturbative corrections to $W$ which do depend on K\"ahler
moduli.  It is proved in \Witten\ that these come entirely from a
class of Euclidean D3 branes wrapping certain 4-cycles $C_{\alpha}$
in $X$. The authors of \Burgess\ discuss the extension of this
nonrenormalization theorem to vacua with nonzero fluxes.\foot{
Strong IR dynamics in gauge theories arising from
D7 branes wrapping 4-cycles
$C$ in $X$ can also contribute to the superpotential for K\"ahler
moduli.  Such contributions are also exponentially suppressed at
large ${\rm vol}(C)$.}

The D3-brane instantons generate a correction to the superpotential
\eqn\corr{\Delta W = \sum_{\alpha} f_{\alpha}(z_a) {\rm Exp}(-{\rm
vol}(C_{\alpha})) + \cdots } where $z_a$ are complex structure and
brane position moduli, $f$ is a one-loop determinant, and $\cdots$
denotes additional terms which are exponentially smaller at large
volume than the terms we kept. The leading terms clearly scale like
$e^{-{R^4\over {(\alpha^\prime)^2}}}$ and vanish exponentially
quickly at large volume. Any masses arising predominantly from
\corr\ will scale as \eqn\mass{m \sim e^{-{R^4\over
{(\alpha^\prime)^2}}} } up to power law prefactors at large $R$. Of
course these masses are parametrically $\ll {1\over R}$.

\subsec{Quantum corrections to $K$}

In models arising from supergravity or string theory, the additional
terms in the 4d potential \potis\ arising from K\"ahler
covariantization of the partial derivatives \eqn\kahcoh{\partial_i W
\to D_i W =
\partial_i W + {K_{,i} \over M_P} W} (as well as from the $e^K$
prefactor in \potis) can play an important role in the effective
potential. Fields which do not receive a mass from the leading order
potentials generated by \eg\ fluxes, are quite plausibly stabilized
in many models by these supergravity corrections to global quantum
field theory.  It is then interesting to ask for the form of such
corrections.

For IIB orientifolds, the leading such corrections were computed in
\bbhl. A nice summary of the resulting potentials appears in \Haack.
For further work on these corrections and their physical
applications see \refs{\kha,\khb,\khc}. The basic structure is the
following. Defining the complex scalar \eqn\rhodef{\rho = \left(\int
C_4\right) ~+~i {R^4\over {(\alpha^\prime)^2}}~} (where the integral
defining ${\rm Re}(\rho)$ is taken over an appropriate 4-cycle and
gives rise to an axion field), one finds \eqn\kahpot{K = - 3 ~{\rm
log} \left( \rho - \overline\rho + f_1(z_a) \right) + {f_2(z_a)\over
{\rho - \overline\rho}} + {f_3(z_a)\over {(\rho
-\overline\rho)^{3/2}}}  + {f_4(z_a)\over {(\rho -\overline\rho)^2}}
+ \cdots.} Again $z_a$ runs over the complex moduli, brane
positions, and the axio-dilaton.

For any fields which have not received a potential from the bulk
fluxes of \S3.1, including the correction to $K$ in the expansion of
the 4d potential \potis\ can yield a mass. Integrating out the
fields which receive a mass from three-form fluxes, \Haack\ finds a
potential which takes the schematic form \eqn\potschem{V = {c_1
\over R^{18}} + {c_2 \over R^{20}} + \cdots,} where $c_{1,2}$ depend
on combinations of the remaining light fields.\foot{In some models
where even non-perturbatively $W$ is independent of $\rho$, these
corrections may suffice to fix $\rho$. This requires competition
between different orders of perturbation theory. To find reliable
vacua of this sort one requires $|c_2| \gg |c_1|, c_2 > 0, c_1 < 0$,
which can plausibly happen in some fraction of models.}

Because of the $R^{-18}$ fall-off of the leading term in \potschem,
any fields which receive their leading mass from such corrections
are parametrically lighter than those of \S3.1, and have masses $\ll
{1\over R}$.

\newsec{Asymmetric models: Heterotic M-theory}

So far, we have assumed that $X$ is fairly isotropic so that it can
be characterized by a single Kaluza-Klein scale $1/R$. In some
cases, it is natural to assume that the compactification manifold
has some dimensions much larger than others.  A frequently discussed
example is the compactification of M-theory on a Calabi-Yau manifold
$X$ of radius $R_X\sim \ell_{11}$ times an interval of length $R\gg
R_X$ separating the two $E_8$ walls \EdG. This higher dimensional
setup could possibly lead to a model of anomaly mediation in
M-theory \RS.

To find the masses of complex structure moduli of $X$ which
receive a mass from $G_4$ flux, one can do an analysis very
similar to that in \S3.1.\  The allowed modes of $G_4$  have a
single leg on the interval. Therefore, the scaling of the potential
arising from G-flux will be \eqn\gscal{V \sim {1\over R^2} \int_{X
\times S^1/Z_2} ~\sqrt{-g} G_{ijkl} g^{im} g^{jn} g^{kp}g^{lq}
G_{mnpq} \sim {1\over R^3}} where the prefactor to the integral
comes from the Weyl rescaling to 4d Einstein frame. Since $M_P^2
\sim R$, the moduli masses arising from fluxes turn out to scale as
\eqn\hetmass{m \sim {1\over R}~.} It then becomes a detailed
numerical question, whether or not the suppression \falloff\ kicks
in for a given complex structure mode.

The K\"ahler moduli of $X$, however, do not receive a mass from G-flux. In
perturbative string theory they could, in some models, develop a
superpotential at the level of worldsheet instantons \dsww. In the
strong coupling regime, the worldsheet instantons become membrane
instantons extended across the interval. Hence the moduli masses
will be exponentially small at large $R$ \eqn\instac{m \sim {\rm
exp}(-R/\ell_{11})~.} K\"ahler potential corrections could potentially
give these moduli masses larger than \instac\ but parametrically
smaller than \hetmass. The discussion of these corrections would be
analogous to the one in \S3.4.

The models commonly discussed in the literature incorporate gauge bundles
$E \to X$, breaking the $E_8$ gauge symmetries, which have $c_1(E) = 0$.
In models involving a gauge bundle with $c_1(E)\neq0$,\foot{See
\eg\ \Sharpe\ for a detailed discussion.}
there is
an additional contribution to some K\"ahler moduli masses from the $E_8$
gauge fluxes.
The potential due to gauge fluxes (with field strength $F$) scales as
\eqn\gapo{V\sim {1\over {R^2 R_X^{12}} }\int_X \sqrt{g_{10}} F_{ij}g^{ik} g^{il}
F_{kl}\sim {1\over {R_X^{10} R^2}}\sim {1\over R^2},} where the factor
${1\over {R^2 R_X^{12}}}$ comes from Weyl rescaling into 4d Einstein frame and,
in the last equality, we set $R_X\sim \ell_{11}.$ Hence, in this class
of models, some K\"ahler moduli get masses \eqn\kasa{m\sim
\ell_{11}^{-1}\sqrt{{\ell_{11}\over R}}~,} which lifts them above the bound
\massbound.\

This is promising.  However,
the potential induced by nontrivial $c_1(E)$
just imposes the geometric condition (required for supersymmetry)
\eqn\chernco{c_1(E) \wedge J \wedge J ~=~0}
where $J$ is the K\"ahler form of $X$ \gsw.
This means that the leading order effect of gauge
fluxes does not induce a potential for the overall volume modulus of
$X$, since any solution of \chernco\ can be scaled up $J \to \lambda J$
while preserving the condition \chernco.
So the overall volume modulus of $X$
should receive its leading mass from some other
source or from a quantum correction to the potential generated by gauge
flux, and will generically have a mass which is parametrically
lighter than $1/R$.

There is also a modulus controlling $R$, the length of the interval.
It enjoys various special properties, as described in \LutySun.
For instance, this field has
flavor-blind couplings to the observable $E_8$, and in fact does not
have dangerous dimension-six couplings at all, at leading order.
It is therefore ${\it not}$
important for this field to satisfy \massbound.
More generally, one could conjecture that generic fields which control
separations of the Standard Model brane from other branes, but do not control
the geometry or field theory
couplings of the Standard Model brane itself, would
enjoy universal
couplings to the different generations and may safely violate \massbound.
In contrast, any bulk
moduli controlling the geometry or couplings of the Standard Model brane should
be required to satisfy \massbound, since they ``know about'' the origin
of flavor and a priori can introduce flavor violations into the SUSY-breaking
soft terms.\foot{We thank R. Sundrum for helpful discussions of this
distinction.}
We conclude that, if
in a given example the subset of K\"ahler modes of $X$ which violate
\massbound\ also enjoy the special properties described in \LutySun,
and if the
numerical factors in \hetmass\ are favorable, one could
potentially find a working model in this asymmetric regime.

We now make some remarks about the extension of this logic to type II
models.
We will discuss IIB, the mirror statements apply in IIA.
In IIB models, the Standard Model is typically realized on some stack
of D-branes.  To leading order, the superpotential couplings are controlled
by complex structure moduli while the gauge couplings and D-terms
depend on K\"ahler moduli \DougAlb.  For example
in the particular D3-brane model of
\Herman, the $dP_8$ blow-up modes control the pattern of gauge
symmetry breaking
due to FI terms,
while the 8 deformations of complex structure control Yukawa
couplings.  These modes
should be required to satisfy \massbound.  For other bulk moduli,
the question would
be more model dependent.  Some subset could enjoy the special properties
discussed in \LutySun.

\newsec{Discussion}

The results of \S3-4\ basically arise because string theory, at
distances large compared to $l_s$, behaves very much like a local
field theory. For instance, the known corrections to the 4d
effective superpotential which do not have a local 10d description
(\eg\ which involve Euclidean brane instantons) are exponentially
small at $R> l_s$. But local contributions to the energy density in
a compactification on $X$ grow at most as fast as ${\rm vol}(X)$.
The Weyl rescaling to go to 4d Einstein frame, which rescales the 4d
effective potential by $({\rm vol}(X))^{-2}$, more than overcomes
this fastest possible growth, yielding potentials which vanish
quickly as $R$ increases. Since the bulk masses come from the
dependence of these potentials on the bulk moduli $\phi_a$, it
becomes very challenging to find models where the bulk moduli have
masses $\geq {1\over R}$. It may be possible to prove more rigorous
bounds in some spacetimes by using the AdS/CFT correspondence
\Cumrun.

Our approach here was to apply simple scaling arguments to various
ingredients which can be invoked in different well-known classes of
string compactifications. If indeed all low-energy supersymmetric
models have some bulk field with mass which is parametrically
smaller than the bound \massbound, it would be nice to find a
conceptual proof of this result. The locality argument above
eliminates many but not all possibilities.  For instance, the
potential energy from zero-form fluxes enjoys sufficiently slow
fall-off  $\sim 1/R^6$ to potentially give high-scale bulk masses.
However, we are not aware of any models which incorporate these in a
way that actually leads to moduli masses satisfying  \massbound.

There are a few obvious caveats to the arguments presented here.
Firstly, we have only discussed parametric scalings.  It may be that
numerical factors, in a regime where $R$ is greater than $l_s$ but
not very large, can sufficiently enhance moduli masses to allow the
Yukawa suppression \falloff\ to kick in.

Secondly, one can imagine scenarios where the number of moduli is
extremely small and couplings of the moduli to the branes are highly
constrained by symmetry arguments.  For example, in \LutySun\ it was
argued that in five dimensional models compactified on an interval,
the overall radion controlling the separation of the ``end of the
world'' branes has highly constrained couplings to the branes, which
are flavor-blind at leading order.  Such modes will only need to
satisfy a weaker bound than \massbound, and hence supersymmetric
string models with very few moduli in the leading approximation
(\eg\ a hypothetical Calabi-Yau space with $h^{1,1}=1$ and
$h^{2,1}=0$) could provide interesting exceptions to our reasoning.

Thirdly, one could always consider highly anisotropic
compactifications. The models of \S4 fall into this class, but other
anisotropic models may be equally or more interesting in this
regard. In the case of gaugino mediation, assuming one wishes to
preserve the successful high-scale grand unification resulting from
logarithmic running of the MSSM couplings \Savas, there is not much
room for anisotropy.  The large extra dimensions should not be
larger than the GUT scale, however the GUT scale is quite close to
the string scale, so one does not want the extra dimensions in which
gauge fields propagate to be substantially larger than
$l_s$.\foot{One could loosen this constraint by either giving up
unification, or by adding extra ingredients to the MSSM to yield
low-scale unification. Neither option is particularly attractive.}
In the case of anomaly mediation, the situation is less constrained.
It is conceivable that consideration of highly anisotropic models,
combined with special arguments about the couplings (to the Standard
Model fields) of the moduli controlling the sizes of the large
dimensions, could yield exceptions to our arguments.

Perhaps more generally, one should keep in mind that the history of
``no-go'' arguments in string theory is somewhat tortured,
suggesting a no-go theorem for no-go theorems.  In addition to the
examples provided in the introduction, there are the famous examples
involving chirality in compactifications of 11d supergravity
\witteno, warped compactifications \warpedno, and many others. In
this spirit, one can view the arguments here as a provocation to
find new classes of string models where large bulk masses can be
achieved and the clever constructions of
\refs{\RS,\Murayama,\Gaugino} can be realized.

\medskip
\centerline{\bf{Acknowledgements}}

We would like to thank S. Dimopoulos, E. Silverstein and J. Wacker
for useful
discussions.  We are especially grateful to R. Sundrum for
several helpful conversations, and
for detailed comments on an early version of
this note.
This work was supported in part by a David and
Lucile Packard Foundation Fellowship for Science and Engineering,
the NSF under grant 0244728, and the DOE under contract
DE-AC02-76SF00515.

\listrefs
\end